\documentclass[]{interact}

\usepackage{graphicx}
\usepackage{dcolumn}
\usepackage{bm}
\usepackage{verbatim}
\usepackage{float}
\usepackage{xcolor}
\usepackage{amsmath}
\usepackage{caption}
\usepackage[numbers,sort&compress,merge]{natbib}
\bibpunct[, ]{[}{]}{,}{n}{,}{,}

\begin{document}

\title{Quantum-State-Specific Reaction Rate Measurements for the Photo-induced Reaction Ca$^+$ + O$_2$ $\rightarrow$ CaO$^+$ + O}

\author{
\name{Philipp C. Schmid,\textsuperscript{a} Mikhail I. Miller,\textsuperscript{a} James Greenberg,\textsuperscript{a} Thanh~L. Nguyen,\textsuperscript{b} John F. Stanton\textsuperscript{b} and H. J. Lewandowski\textsuperscript{a} }
\affil{\textsuperscript{a}JILA and the Department of Physics, University of Colorado, Boulder, Colorado 80309-0440, USA; \textsuperscript{b}Quantum Theory Project, Departments of Chemistry and Physics, University of Florida, Gainesville, Florida 32611, USA}
}

\maketitle

\begin{abstract}
	Atoms and molecules often react at different rates depending on their internal quantum states.  Thus, controlling which internal states are populated can be used to manipulate the reactivity and can lead to a more detailed understanding of reaction mechanisms. We demonstrate this control of reactions by studying the excited state reaction reaction Ca$^+$ + O$_2$ $\rightarrow$ CaO$^+$ + O.  This reaction is exothermic only if Ca$^+$ is in one of its excited electronic states. Using laser-cooling and electrodynamic trapping, we cool and trap Ca$^+$ at millikevin temperatures for several minutes. We can then change the fraction of time they spend in each of the two excited states by adjusting the detunings of the cooling lasers. This allows us to disentangle the reactions that begin with Ca$^+$ in the $^2$P$_{1/2}$-state from the ones where Ca$^+$ is in the $^2$D$_{3/2}$-state. Using time-of-flight mass spectrometry, we determine independent reaction rate constants for Ca$^+$ in both electronically excited quantum states. 
	
\end{abstract}

\begin{keywords}
linear ion trap; Coulomb crystals; chemical reactions; photo-assisted reaction
\end{keywords}

\section{Introduction}

The study of cold, controlled molecular reactions and collisions in the gas phase is a steadily growing field, offering many opportunities for detailed studies of interactions \cite{Dulieu2009,Dulieu2018,Carr2009,doyle2018}. Several different experimental pathways are currently used to explore cold molecular interactions with neutral molecules. These include, but are not limited to, crossed-beam experiments using a Stark decelerator \cite{Vogels2015,bas2018}, merged-beam experiments \cite{Klein2017,Gordon2017}, co-trapped atoms and moleucles \cite{Parazzoli2011,Akerman2017}, buffer-gas cooled molecules inside magnetic traps \cite{Weinstein1998}, trapped Rydberg molecules \cite{Allmendinger2016}, laser-cooled molecules  \cite{Anderegg2017,Kozyryev2017,Norrgrd2016,Williams2017,Hummon2013} and ultracold bi-alkali atom molecules \cite{Kraemer2006,Ospelkaus853}. Another set of experiments uses laser-cooled atomic ions confined in rf traps as reactants, or as a cold bath to sympathetically cool molecular ions for reaction studies \cite{Willitsch2012,Heazlewood2015}.

While most of these experiments benefit from reducing the translational motion via cooling to temperatures below T$\leq$1 K, controlling the internal states of the reactants can offer additional insights into understanding the reactions. Preparing the reactants in a single quantum state could elucidate the state's influence on reaction pathways. There have been some experiments that explore cold reactions where the internal states of the atomic reactant have been controlled including, but not limited to, Ca$^+$ + O$_2$ \cite{Drewsen2002}, Ca$^+$ + H$_2$ \cite{Kimura2011}, Ca$^+$ + H$_2$O \cite{Okada2003}, Ba$^+$ + $^{87}$Rb \cite{Hall2013}, or Ca$^+$ + (CH$_3$F, CH$_2$F$_2$, or CH$_3$Cl) \cite{Gingell2010}, and Be$^+$ + H$_2$O \cite{Yang2018}. Theses experiments take advantage of the control of electronic states of the atomic ion using cooling lasers.  However, preparation of pure state-selective molecular samples is significantly more challenging \cite{Tong2012,Chang2013,Allmendinger2016}.

Here, we explore the reaction between laser-cooled Ca$^+$ ions and neutral O$_2$, where the reaction can proceed only when Ca$^+$ is in one of two excited electronic states. We control the relative populations in the different quantum states by the wavelengths of the cooling lasers. Although the reaction of Ca$^+$ + O$_2$ has been observed previously \cite{Drewsen2002}, no details on the reaction kinetics or the influence of the quantum states of Ca$^+$ have been reported. In this work, we present a detailed study on the reaction Ca$^+$ + O$_2$ $\rightarrow$ CaO$^+$ + O, including the absolute reaction rate constants for reactions starting with Ca$^+$ in either the $^2$P$_{1/2}$-state or $^2$D$_{3/2}$-state.

In this paper, we begin by discussing the energetics of the reaction (Sec. \ref{sec:cao2}) before discussing the experimental setup (Sec. \ref{sec:apparatus}). Next, we present results of the reaction measurements and the dependence of the rates on the quantum state of Ca$^+$ (Sec. \ref{sec:rate}). We then discuss the implications of our results (Sec. \ref{sec:discussion}), and conclude by giving an outlook for future experiments (Sec. \ref{sec:summary}).

\section{\label{sec:cao2}C\lowercase{a}$^+$ + O$_2$ Reactions}

\begin{figure}[H]
\begin{center}
\includegraphics[width=\columnwidth]{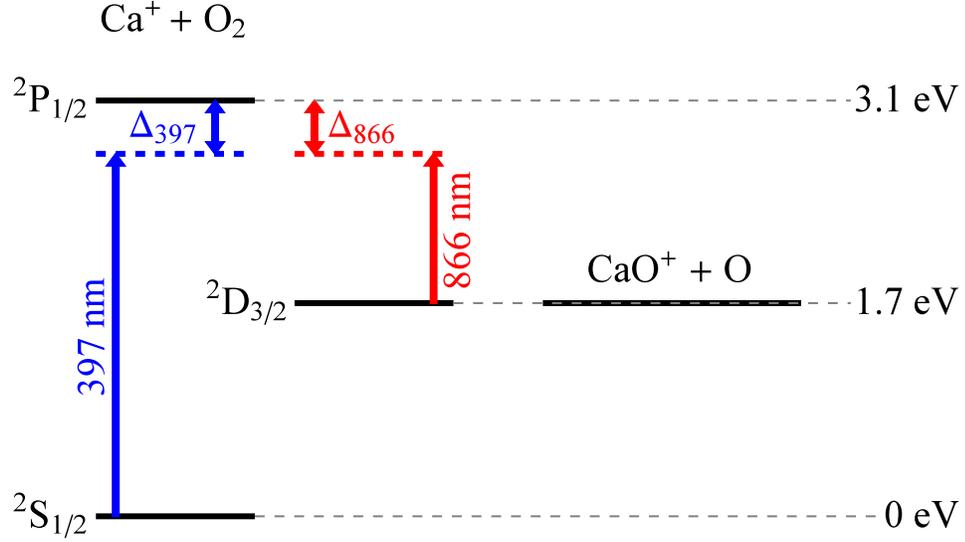}%
\caption{\label{fig:fig1}Diagram of the Ca$^+$ energy structure and product channel energy (energies are not to scale). The main laser cooling transition for Ca$^+$ at 397 nm excites the ion to the $^2$P$_{1/2}$-state. The ion can decay to the ground $^2$S$_{1/2}$- or $^2$D$_{3/2}$-state. A second laser at 866 nm repumps population that accumulates in the dark $^2$D$_{3/2}$-state back into the cooling cycle. Here, $\Delta_{397}$ and $\Delta_{866}$ represent the detuning of the cooling lasers with respect to the two cooling transition frequencies. The reaction with O$_2$ is energetically possible in either of the two excited states, but not in the ground state. }
\end{center}
\end{figure}

In the experiments presented here, we show measurements of the reactions between Ca$^+$ ions, which have been laser-cooled to millikelvin temperatures, and O$_2$ molecules at room temperature.  This leads to an average collision energy of $\sim$  160 K.  The reaction of Ca$^+$ with O$_2$ is highly endothermic (+ 1.7 eV) when Ca$^+$ is in the ground $^2$S$_{1/2}$-state, but exothermic when the ion is in either the $^2$P$_{1/2}$- or $^2$D$_{3/2}$-states, which are populated during the laser cooling process (Fig. \ref{fig:fig1}). Therefore, the two reaction pathways that are exothermic are:

\begin{equation}
Ca^+ (^2P_{1/2}) + O_2 \rightarrow CaO^+ + O \quad \Delta{}E = -1.45(2)~eV
\label{eqn:eqn1}
\end{equation}

\begin{equation}
Ca^+ (^2D_{3/2}) + O_2 \rightarrow CaO^+ + O \quad \Delta{}E = -0.02(2)~eV
\label{eqn:eqn2}
\end{equation}
We note that the exothermicity of the $^2$D$_{3/2}$-state reaction is equivalent to the uncertainty in the calculated energy. However, the collision energy is observed to be sufficient to counter any endothermicity, if present. Thus, the reaction rate for Ca$^+$ + O$_2$ will depend on the population of both excited states of Ca$^+$.

\begin{figure}[H]
\begin{center}
\includegraphics[width=\columnwidth]{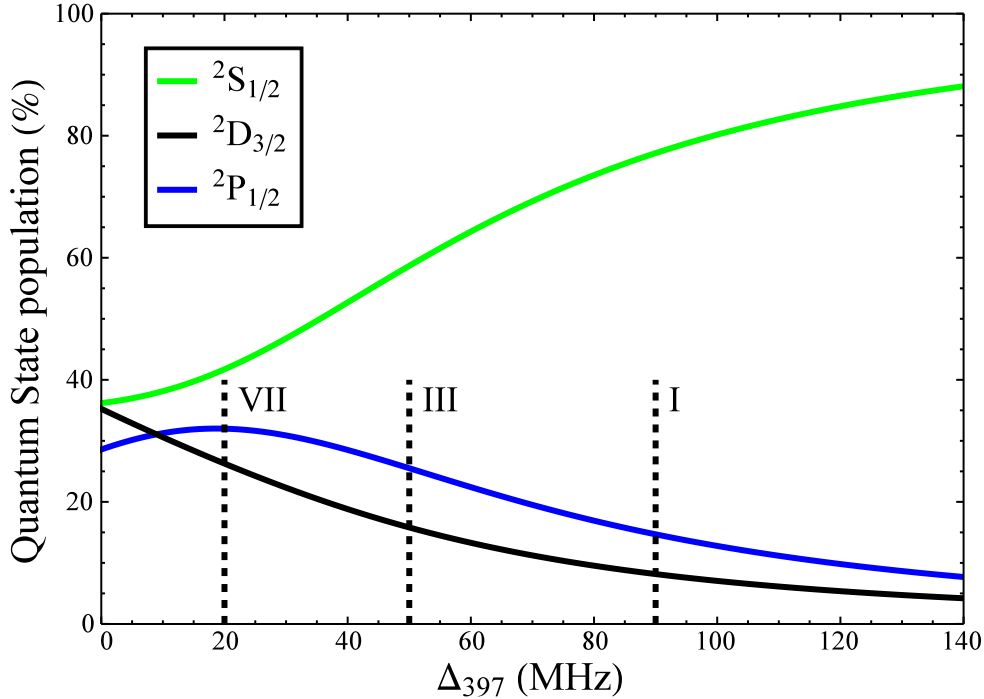}%
\caption{\label{fig:fig2}Predictions of the state distribution of the Ca$^+$ ion from the 3-level OBE with a fixed 866~nm detuning (-35 MHz). The state populations for the $^2$S$_{1/2}$, $^2$P$_{1/2}$, and $^2$D$_{3/2}$ states are shown. The vertical lines represent the parameters used for three measurements used to determine the rate constants presented in Section \ref{sec:rate} (see Table \ref{tab:tab2}). To obtain the other state distributions shown in Table \ref{tab:tab2}, we also changed the 886~nm laser detuning. 
}
\end{center}
\end{figure}

We control the relative populations of Ca$^+$ in its two excited states by adjusting the detuning of both the main cooling laser at 397 nm and the repump laser at 866 nm (Fig. \ref{fig:fig1}). To determine the fraction of the time the Ca$^+$ spends in each of its three electronic states, we use the 3-level optical Bloch equations (OBE) calibrated to the magnitude of the fluorescence from the Coulomb crystal following the method described in \cite{Gingell2010}. Once calibrated, the 3-level OBE are used to predict the populations in each state based on the detunings of both lasers used for laser cooling. An example set of curves for the three states is shown in  Fig. \ref{fig:fig2}. Here, the repump laser detuning is fixed, while the main cooling laser detuning is varied. 

\section{\label{sec:apparatus}Ion Trap Apparatus}
Details of the experimental setup for measuring reactions between cations and neutrals have been described previously \cite{Schmid2017,Greenberg2018}. In the following, we present a short description of the apparatus with an emphasis on aspects relevant for the present experiment.

Ca$^+$ cations are created from non-resonant photo-ionization at 355 nm (7 mJ/pulse, 10 Hz, beam diameter $\approx$ 1 mm$^2$) and trapped using  a segmented, linear quadrupole ion trap \cite{Schmid2017}. The trapped Ca$^+$ ions are laser cooled \cite{Eschner2003} to secular temperatures close to the Doppler cooling limit resulting in the formation of Coulomb crystals \cite{Drewsen2015}. Laser cooling is performed on the $^2$S$_{1/2}$- $^2$P$_{1/2}$ transition, driven at 397 nm by a fiber-coupled diode laser (NewFocus, 3.5 mW, 600 $\mu m$ beam diameter). A second fiber-coupled diode laser at 866 nm (NewFocus, 9 mW, 2 mm beam diameter) is used to repump the ions back into the cooling cycle on the $^2$P$_{1/2}$ - $^2$D$_{3/2}$ transition. Both laser frequencies are measured and locked using a wavemeter (HighFinesse/ {\AA}ngstrom WSU-30) with a precision of $\pm$ 2.5 MHz around the center frequency. The wavemeter is calibrated daily to a 780 nm laser, locked to a transition in $^{87}$Rb by saturation absorption spectroscopy. 

Detection of the trapped ions is done by two different methods. We can image the fluorescence from the Coulomb crystal onto an intensified CCD camera or eject the ions into a time-of-flight mass spectrometer (TOF-MS), which is radially coupled to the linear ion trap \cite{Schmid2017}. Optical detection has the advantage of being non-destructive, but is sensitive to only Ca$^+$. Non-fluorescing ions can be inferred only by a change in the shape of the Coulomb crystal in combination with results from molecular dynamics simulations \cite{Okada2015b}. In the present experiment, we use the optical detection of pure Ca$^+$ crystals to determine the initial number of trapped atomic ions by fitting the area of the crystal and knowing the density of ions in the trap \cite{Schmid2017}. To accurately determine the number of all trapped reactant and product ions, we eject all ions in the trap into the TOF-MS. This allows us to determine the absolute number of ions at each mass in a single shot \cite{Schmid2017}. Thus, we can study the reactions by the increase of products, as well as by the loss of reactants.

An experimental sequence starts by loading a Ca$^+$ Coulomb crystal into the ion trap and removing any impurity ions by varying the trapping fields. Next, an image of the crystal is taken to determine the initial number of Ca$^+$ ions. At this point, O$_2$ is introduced into the chamber through a leak valve for a predetermined length of time.  After the set reaction time has been reached, all trapped ions are ejected into the TOF-MS and a mass spectrum is produced. The total number of ions (product + reactant) is also determined at each time point to ensure the number of ions remains constant during a reaction measurement and ions are not lost from the trap. This check is especially important when measuring the highly exothermic, excited-state reaction of Ca$^+$ + O$_2$, where the charged products may have $>$ 1~eV of kinetic energy.

There is one notable improvement to the apparatus compared to our previous work \cite{Schmid2017}, which concerns how we introduce the neutral reactant. Instead of having the O$_2$ present during the trap loading sequence, we now implement a pulsed-leak valve~\cite{Jiao1996}. The valve consists of an all-metal leak valve (Kurt Lesker LVM series) with a 3-way solenoid valve (Parker Hannifin) connected to the input port. The 3-way valve connects the input port of the leak valve to either a roughing pump or a bottle of O$_2$ diluted to 7\% in argon. In this configuration, the leak valve is held open at a desired leak rate at all times and the 3-way valve controls whether or not O$_2$ flows into the chamber. Using the 3-way valve to initiate the introduction of O$_2$ in to the trapping chamber, we achieve a stable background O$_2$ pressure in around 1$\textendash$2 seconds. This time sets the uncertainty in the initial time for the reactions.

\section{\label{sec:rate}Reaction Rate Measurements}
We model the reactions in our experiment using pseudo-first-order kinetics, where we assume that the density of O$_2$ does not change during a reaction measurement.  We model the system by including reactions that start with the Ca$^+$ in the two excited states. Equation (\ref{eqn:eqn3}) describes the growth of the charged product, CaO$^+$.
\begin{equation}
[CaO^+] = [Ca^+_0](1-e^{-k_{eff}[O_2]t})
\label{eqn:eqn3}
\end{equation}
Here, [CaO$^+$] is the number of CaO$^+$ ions as a function of reaction time, t, [Ca$^+_0$] is the initial number of trapped Ca$^+$ ions, $k_{eff}$ is the effective reaction rate constant, and [O$_2$] is the oxygen concentration in the chamber. The effective reaction rate constant, k$_{eff}$, is given by
\begin{equation}
k_{eff} = k_{p}f_p + k_{d}f_d,
\label{eqn:eqn4}
\end{equation}
where k$_{p/d}$ is rate constant for reactions with Ca$^+$ in the P/D-state, and f$_{p/d}$ is the fraction of the Ca$^+$ ions in the P/D-state. 

\begin{figure}[H]
\begin{center}
\includegraphics[width=\columnwidth]{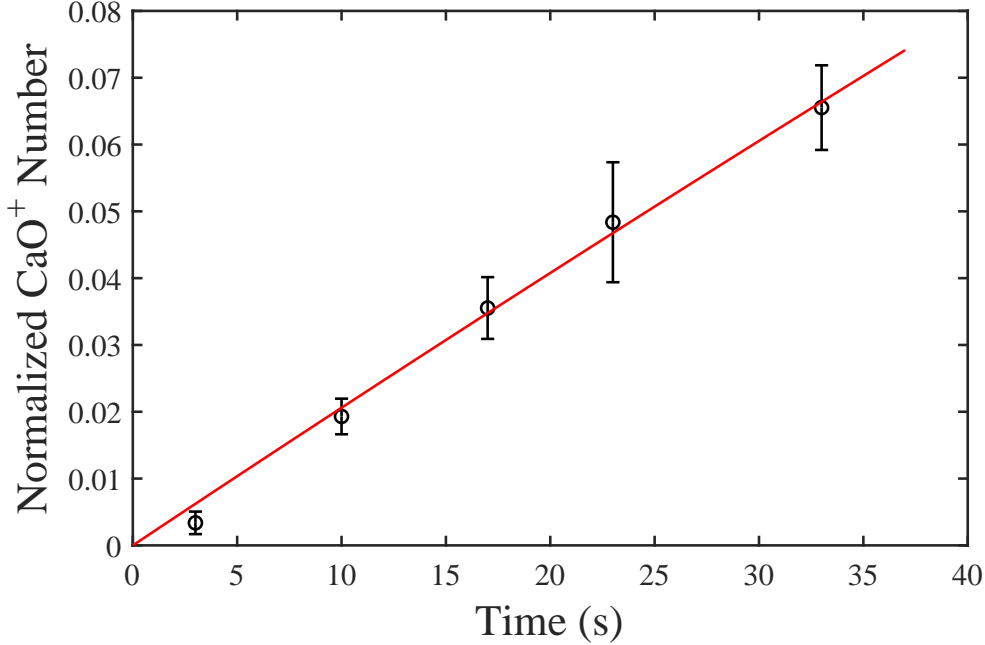}%
\caption{\label{fig:fig3}Example reaction measurement showing the growth of CaO$^+$ product ions, normalized by the initial number of Ca$^+$ ions ($\sim$ 900).  The fraction in the $^2$P$_{1/2}$-state is 0.22 and $^2$D$_{3/2}$-state is 0.36, with an O$_2$ density of $3.5\times 10^{7} cm^{-3}$.  The solid line is a fit using Eqn. (\ref{eqn:eqn3}), where k$_{eff}$ is the only free parameter. The error bars correspond to the standard error of the mean of 5 measurements at each time point.}
\end{center}
\end{figure}

\begin{figure}[H]
\begin{center}
\includegraphics[width=\columnwidth]{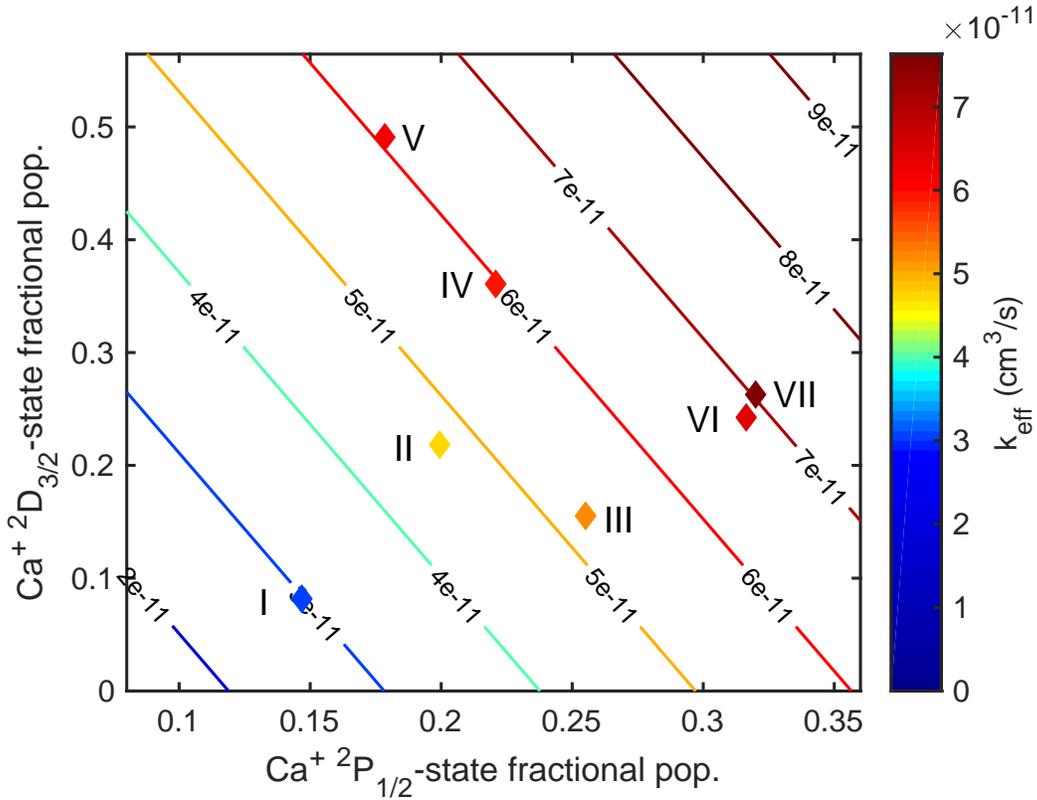}%
\caption{\label{fig:fig4}Dependence of the effective rate constant, k$_{eff}$, on the $^2$P$_{1/2}$- and $^2$D$_{3/2}$-state populations of Ca$^+$. The experimental measurements are shown as diamonds, where the color of the point represents the value of k$_{eff}$. For all data shown, the O$_2$ concentration was kept constant at $3.5\times 10^{7} cm^{-3}$.  We perform a two-dimensional fit to the data using  Eqn. \ref{eqn:eqn4}, which allows us to extract $k_p$ and $k_d$. The contour lines represent constant k$_{eff}$ values of the fit.}
\end{center}
\end{figure}

To extract the quantum-state dependent reaction rate constants, we first measure k$_{eff}$ at several different state populations.  We accomplish this by varying the detunings of the cooling lasers, while maintaining a constant O$_2$ concentration. An example of one such rate measurement is shown in Fig. \ref{fig:fig3}. With our experimental parameters, the $^2$P$_{1/2}$-state population can be varied between 14\% and 35\%, while the $^2$D$_{3/2}$-state range is between 50\% and 14\%. We are not able to work outside this range, as either the heating rates become too large such that we can not maintain a stable Coulomb crystal due to the energy imparted by collisions with the background gas or the population is not possible due to the coupling in a three-level system.  A summary of the determined k$_{eff}$ for the different state populations is given in Table \ref{tab:tab2}. We perform a two-dimensional fit to these data using Eqn. \ref{eqn:eqn4} as the model to extract two fit parameters, k$_p$ and k$_d$. The data and the fit can be seen Fig. \ref{fig:fig4}. The measurements are shown as diamond points, where the color represents k$_{eff}$.  The fit is shown as contour lines of constant k$_{eff}$. The rate constants from the fit are given in Table \ref{tab:tab3}. The rate constant for Ca$^+$ in its $^2$D$_{3/2}$-state is about three times slower than $^2$P$_{1/2}$-state reactions.

\begin{table}[H]
\caption{\label{tab:tab2}Measured reaction rate constants for pairs of Ca$^+$ excited state populations shown in Fig. \ref{fig:fig4}. The fractional populations in the $^2$P$_{1/2}$-state ($f_P$) and $^2$D$_{3/2}$-state ($f_D$) are shown with the corresponding measured effective rate constant (k$_{eff}$).  The data point numbers correspond to measurements shown in Fig. \ref{fig:fig4}.} 
\begin{center}
\begin{tabular}{c|c|c|c}
Data point & ~~~$f_P$~~~ & ~~~$f_D$~~~ & k$_{eff}$ (10$^{-11}$ cm$^3$ s$^{-1}$)\\
\hline\hline
I & 0.15 & 0.08 & 3.3\\
II & 0.20 & 0.22 & 4.6\\
III & 0.26 & 0.16 & 5.1\\
IV & 0.22 & 0.36 & 6.0\\
V & 0.18 & 0.49 & 6.1\\
VI & 0.32 & 0.24 & 6.3\\
VII & 0.32 & 0.26 & 7.6\\
\end{tabular}
\end{center}
\end{table}

\begin{table}[H]
\caption{\label{tab:tab3}The resulting rate constants k$_{p/d}$ for the three quantum states of Ca$^+$. Uncertainties are from the 90\% confidence interval for the fit parameters, k$_{p/d}$.}
\begin{center}
\begin{tabular}{c|c}
Ca$^+$ state & k (10$^{-10}$cm$^3$ s$^{-1}$)\\
\hline\hline
$^2S_{1/2}$ & 0 \\
$^2P_{1/2}$ & 1.7(1)\\
$^2D_{3/2}$ & 0.6(1)\\
\end{tabular}
\end{center}
\end{table}

\section{\label{sec:discussion}Discussion}

There are two possible explanations as to why the rate constants for the different quantum states are not equivalent. The first of these is related to the energy of the $^2$D$_{3/2}$-state. The calculated energy of the Ca$^+$($^2$D$_{3/2})$ + O$_2$ entrance channel is equivalent, to within the estimated uncertainty, to the CaO$^+$ +  O outgoing channel. If the reaction is actually endothermic by an amount comparable to the collision energy, then only a portion of the Boltzmann distribution of the O$_2$ have enough energy to react. Assuming the ions are at rest and the O$_2$ is at 300 K, the estimated collision energy is 160 K. To have a reduction of the rate constant by a factor of three (i.e., only one third of the collision have enough energy to react), an energy barrier of ca. 15 meV would have to exist. This is reasonable considering the uncertainty in the calculated energies. 

\begin{figure}[H]
\begin{center}
\includegraphics[width=\columnwidth]{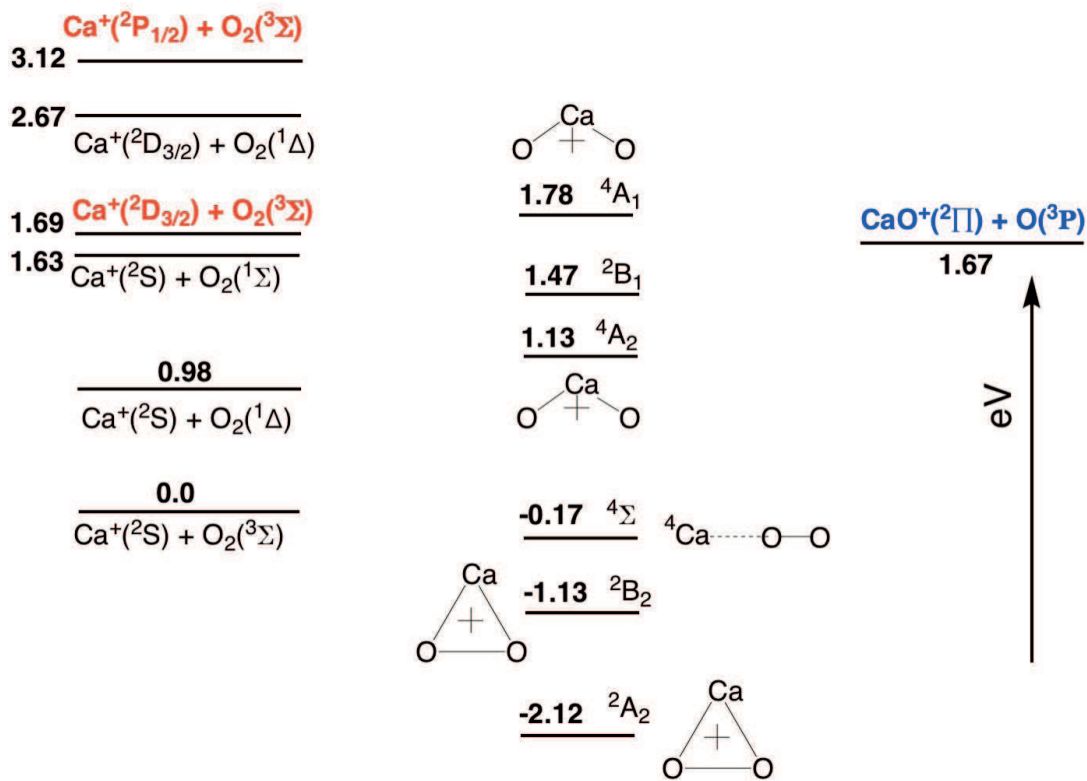}%
\caption{\label{fig:fig5}An energy-level diagram featuring electronic states of reactants, intermediates, and the observed CaO$^+$ + O($^3$P) product channel.}
\end{center}
\end{figure}

The other possible explanation involves considering the intermediate states in the reaction. To help guide understanding of the experimental results, quantum chemical calculations were carried out to locate potentially relevant stationary points on multiple electronic state potential energy surfaces (PES). A coupled-cluster method with single, double, and perturbative triple excitations (CCSD(T)) \cite{Bartlett1990,Raghavachari1989} in combination with Dunning's cc-pCVTZ basis set \cite{Dunning1989} was used to fully optimize all geometries. The same level of theory was then used to compute harmonic vibrational frequencies in order to obtain zero-point vibrational energies. We are able to characterize six different, low-lying intermediate structures of [CaO$_2$]$^+$ including three doublet and three quartet electronic states, each of which is the lowest state of a given symmetry and spin in C$_{2v}$ symmetry, Fig. \ref{fig:fig5}. 

In this work, we have mainly focused on the two reactions relevant to the current experiment, namely Ca$^+$($^2$P$_{1/2}$) + O$_2$ and Ca$^+$($^2$D$_{3/2}$) + O$_2$. As seen Fig. \ref{fig:fig5}, the combination of the excited Ca$^+$ ions with O$_2$ could proceed via a number of electronic states of a charged intermediate.  The doublet channels (which involve a pairing of electrons) are attractive, while the quartet channels likely involve substantial barriers and are closed under the experimental conditions.

Figure \ref{fig:fig5} shows that the association of Ca$^+$ and O$_2$ can potentially lead to several electronic states that lie well below the association asymptote. Favorable combination of the $^2$P$_{1/2}$- and $^2$D$_{3/2}$-states of the atomic calcium ion with ground state O$_2$ correlates with $^2$A$_2$ and $^2$B$_2$ states of the doublet C$_{2v}$ intermediate, where the former arises from pairing of in-plane electrons ($\sigma$ bond) and the latter from a $\pi$ interaction. The lowest states of these symmetries will correlate adiabatically with Ca$^+$($^2$S) + O$_2$($^3\Sigma$), with higher-lying $^2$A$_2$ and $^2$B$_2$ states correlating adiabatically with the Ca$^+$($^2$P$_{1/2}$) or $^2$D$_{3/2}$ + O$_2$($^3\Sigma$) reactants. The minimum-energy path for the approach of the $^2$P$_{1/2}$  state will be in C$_s$ symmetry on excited $^2$A' and $^2$A'' surfaces that correlate with $^2$B$_2$ and $^2$A$_2$ states of the symmetric structure, respectively; the approach of the $^2$D$_{3/2}$-state atomic cation is symmetry allowed in C$_{2v}$.  Although these excited states that adiabatically correlate to reactants have not been explored here, it is likely that internal conversion will ultimately produce vibrationally hot complexes in the lowest two electronic states (Fig. \ref{fig:fig5}), which can then dissociate to the observed products which lie ca. 1.4 eV below the association asymptote (or, alternatively, to ground state calcium ion and O$_2$ in both the $^3\Sigma$ and $^1\Delta$ states). Because of these competing dissociation channels, the rate of the process producing CaO$^+$ + O($^3$P) would potentially be reduced significantly with respect to the rate of intermediate complex formation. Dissociation of the complex to CaO$^+$ + O($^3$P) would certainly encounter a barrier, but the complex will form with roughly 4-5 eV of internal energy, which should be sufficient to surmount it. In addition, another state of the complex that might be involved is $^2$B$_1$, which can dissociate without a barrier to products. It could possibly be formed by nonadiabatic internal conversion along the approach of reactants in the favored $^2$A'' state.   In the event that the barrier to dissociation of the lowest $^2$A$_2$ and $^2$B$_2$ states to the product is above the Ca$^+$($^2$D$_{3/2}$) + O$_2$($^3\Sigma$) entrance channel, the $^2$B$_1$ state could provide a mechanism for the latter to react, since the loss of a ground state oxygen atom from the $^2$B$_1$ state is barrierless.

Note that the above is strictly a thermodynamic plausibility argument; a detailed theoretical study of the kinetics of this process, while perhaps desirable, would require extensive exploration of several excited state potential energy surfaces and then consideration of nonadiabatic dynamics. This is left to future study.

Finally, we compare the total reaction rate constant to the one from the Langivan model to examine the overall reaction rate. We note that a Langevin rate constant is not a good model for this reaction, as it involves excited states and possibly an endothermic reaction, neither of which is taken into account for this model. 

The Langevin rate constant, k$_L$, for Ca$^+$ + O$_2$  is calculated using Egn. \ref{eqn:eqn5}.
\begin{equation}
k_L = \sqrt{\frac{q^2\alpha}{4\mu\epsilon^2_0}},
\label{eqn:eqn5}
\end{equation} 
where q is the charge, $\mu$ the reduced mass of the system, and $\alpha$ the polarizability of the neutral reactant. For the polarizability of O$_2$ of 1.57$\times$10$^{-24}$~cm$^3$~\cite{Rumble2017}, we calculate the Langevin rate constant to be 7 $\times$10$^{-10}$~cm$^{3}$s$^{-1}$. The measured total reaction rate constant, which is the sum of k$_p$ and k$_d$, is k$_{tot}$ = 2.3(1)$\times$10$^{-10}$ cm$^3$s$^{-1}$. These two rate constants are of the same order of magnitude, which is reasonable agreement considering the limitations of the model for describing excited-state reactions. Additionally, we note that the pressure of O$_2$ was measured using an ion gauge, which can give readings that are accurate to within only a factor of 2-3.

\section{\label{sec:summary}Summary}
We measured the products from the photo-induced reaction of Ca$^+$ + O$_2$ $\rightarrow$ CaO$^+$ + O using laser-cooled and trapped samples of Ca$^+$ ions. The product ions were trapped and sympathetically cooled by the atomic ions and detected using a TOF-MS. The reaction is energetically allowed only when the Ca$^+$ is in one of two electronically excited states ($^2$P$_{1/2}$ or $^2$D$_{3/2}$). By adjusting the detunings of the cooling lasers, the quantum-state population of the Ca$^+$ can be controlled. Using this control, we were able to determine reaction rate coefficients for reactions starting in either the $^2$P$_{1/2}$- or $^2$D$_{3/2}$-state. The $^2$P$_{1/2}$-state rate constant is a factor of three larger than for the $^2$D$_{3/2}$-state. We propose two possible explanations for this difference. The $^2$D$_{3/2}$-state reaction might be endothermic by a very small amount, so its reactivity is suppressed with 160 K oxygen, and perhaps proceeds through a different mechanism involving the $^{2}$B$_{1}$ state of the charged CaO$_2$ intermediate.

In future work, we plan to extend these types of measurements to more complex systems, such as reactions with organic molecules, and with additional control over the neutral reactant using a Stark decelerator. We will attach a traveling-wave Stark decelerator \cite{Fabrikant2014,Shyur2018} to the ion trap apparatus, similar to the experiments of the Oxford group \cite{Oldham2014} and the recent proposal by the Willitsch group \cite{Eberle2015}. This will allow us to tune the velocity of the neutral reactant down to 10 ms$^{-1}$, enabling studies of ion-molecule reactions below 10 K. A Stark decelerator can also produce molecules in a single quantum state, thus enabling studies of the influence of the internal molecular states on the reaction. Ultimately, by introducing state-selective molecular ions into the trap \cite{Vitanov2017,Barnum2018}, fully quantum-state resolved molecular reactions at cold temperatures will be possible.

\section*{Funding}
This work was supported by the National Science Foundation (PHY-1734006 
and CHE-1464997), AFOSR (FA9550-16-1-0117), and NIST.

\bibliographystyle{aipnum4-1}
%

\end{document}